\documentclass[sigconf]{acmart}

\AtBeginDocument{%
  }

\copyrightyear{2026}
\acmYear{2026}
\setcopyright{cc}
\setcctype{by}
\acmConference[CIKM '26]{Proceedings of the 35th ACM International Conference on Information and Knowledge Management}{November 07--11, 2026}{Rome, Italy}
\acmBooktitle{Proceedings of the 35th ACM International Conference on Information and Knowledge Management (CIKM '26), November 07--11, 2026, Rome, Italy}
\acmDOI{10.1145/3799682.3841071}
\acmISBN{979-8-4007-2539-5/2026/11}

\usepackage{amsmath}
\usepackage{booktabs}
\usepackage{graphicx}
\usepackage{multirow}
\usepackage{algorithm}
\usepackage{algorithmic}
\usepackage[table]{xcolor}
\usepackage{tcolorbox}
\tcbuselibrary{skins}

\newcommand{\company}{Airbnb}

\newcommand{\brandnew}[1]{#1}
\newcommand{\brandnewblock}{}
\newcommand{\brandendblock}{}

\begin{document}

\title{CASTLE: Contrastive and Seed-Guided Training for Cold-Start Natural Language Search}

\author{Wendy Ran Wei}
\orcid{0009-0008-6259-359X}
\affiliation{\institution{Airbnb}\city{San Francisco}\state{CA}\country{USA}}
\email{wendy.wei@airbnb.com}

\author{Hao Li}
\orcid{0009-0003-8957-9089}
\affiliation{\institution{Airbnb}\city{San Francisco}\state{CA}\country{USA}}
\email{hao.li@airbnb.com}

\author{Weiwei Guo}
\orcid{0009-0008-6848-321X}
\affiliation{\institution{Airbnb}\city{San Francisco}\state{CA}\country{USA}}
\email{weiwei.guo@airbnb.com}

\author{Xiaowei Liu}
\orcid{0000-0002-8797-075X}
\affiliation{\institution{Airbnb}\city{San Francisco}\state{CA}\country{USA}}
\email{xiaowei.liu@airbnb.com}

\author{Xueyin Chen}
\orcid{0009-0009-7339-3849}
\affiliation{\institution{Airbnb}\city{San Francisco}\state{CA}\country{USA}}
\email{s.chen@airbnb.com}

\author{Chun How Tan}
\orcid{0000-0002-2316-1259}
\affiliation{\institution{Airbnb}\city{San Francisco}\state{CA}\country{USA}}
\email{chunhow.tan@airbnb.com}

\author{Malay Haldar}
\orcid{0009-0005-5128-0254}
\affiliation{\institution{Airbnb}\city{San Francisco}\state{CA}\country{USA}}
\email{malay.haldar@airbnb.com}

\author{Soumyadip Banerjee}
\orcid{0009-0004-5508-1861}
\affiliation{\institution{Airbnb}\city{San Francisco}\state{CA}\country{USA}}
\email{soumyadip.banerjee@airbnb.com}

\author{Kedar Bellare}
\orcid{0009-0004-6827-8719}
\affiliation{\institution{Airbnb}\city{San Francisco}\state{CA}\country{USA}}
\email{kedar.bellare@airbnb.com}

\author{Huiji Gao}
\orcid{0009-0006-0424-248X}
\affiliation{\institution{Airbnb}\city{San Francisco}\state{CA}\country{USA}}
\email{huiji.gao@airbnb.com}

\author{Stephanie Moyerman}
\orcid{0000-0003-0526-1077}
\affiliation{\institution{Airbnb}\city{San Francisco}\state{CA}\country{USA}}
\email{stephanie.moyerman@airbnb.com}

\author{Sanjeev Katariya}
\orcid{0009-0008-1519-174X}
\affiliation{\institution{Airbnb}\city{San Francisco}\state{CA}\country{USA}}
\email{sanjeev.katariya@airbnb.com}

\renewcommand{\shortauthors}{Wei et al.}

\begin{abstract}
Deploying natural language search systems presents a critical cold-start
challenge: no real user queries to learn linguistic patterns, and no relevance
labels to train ranking models. We present \brandnew{\textbf{CASTLE} (\textbf{C}ontrastive \textbf{A}nd
\textbf{S}eed-guided \textbf{T}raining for natural \textbf{L}anguage s\textbf{E}arch)},
an LLM-based framework for generating synthetic queries and relevance labels from structured
catalog data, powering \company{}'s natural language search across its full lifecycle.

CASTLE makes three contributions. First, we generate realistic queries by
combining structure-guided prompting with seed queries from user research,
using template, few-shot, and attribute-grounded prompt variants together with
explicit variety mechanisms to prevent query collapse. Second, we produce
relevance labels \emph{by construction} via contrastive listing pairs derived
from booking sessions---achieving near-zero false positives without LLM judgment.
Third, CASTLE's structured input design is flexible: incorporating richer
signals---such as guest reviews and photo captions---alongside listing
attributes enables generation of niche, long-tail queries that reflect
subjective user preferences (e.g.\ ``cozy cabin with fireplace'') beyond what
catalog attributes alone can express.

\brandnew{Compared against InPars-style, Promptagator, and contrastive-only
baselines, CASTLE achieves KL 1.01 vs.\ real users---a 9.2$\times$
improvement over the best baseline (9.33)---and the lowest attribute-type KL
divergence (0.08), outperforming even survey seed queries (0.09). A human evaluation on 200 sampled triplets confirms label quality:
annotators agree with CASTLE labels at 91--93\%.} We deploy production pipelines generating
synthetic examples daily for embedding-based retrieval and ranking evaluation.
\brandnew{Synthetic data remains valuable \emph{beyond cold-start}: it targets tail
queries underrepresented in organic traffic and extends naturally to multi-turn
conversational search.}
\end{abstract}

\ccsdesc[500]{Information systems~Information retrieval}
\ccsdesc[500]{Information systems~Retrieval models and ranking}
\ccsdesc[500]{Computing methodologies~Natural language generation}
\ccsdesc[300]{Computing methodologies~Language resources}

\keywords{synthetic data generation, large language models, cold-start problem,
natural language search, query generation, learning to rank, contrastive
learning, information retrieval, e-commerce search}

\maketitle

\section{Introduction}

Online travel marketplaces have long relied on structured search filters
(destination, dates, guest count, amenities) to match users to listings. At
\company{}, search ranking has been a sustained focus of
optimization~\cite{haldar2019applying,grbovic2018real,haldar2024maps,haldar2025beyond},
with models trained on user interactions to predict booking likelihood. While
precise, filters cannot express open-ended preferences such as ``a quiet retreat
for working remotely.'' The emergence of large language models (LLMs) has made
natural language search technically feasible, but deploying it on a live platform presents a fundamental
\emph{cold-start problem}: without historical user data, there is no signal to
train or evaluate ranking models.

Prior work on search cold-start falls into three categories, each with
limitations. \emph{Rule-based systems} can
bootstrap initial ranking but cannot capture nuanced linguistic intent.
\emph{Human labeling} produces high-quality data but is expensive and slow at
the scale required for modern neural ranking. \emph{Transfer learning from
related domains}~\cite{karpukhin2020dense,ma2021zero} is hampered by domain shift between
passage retrieval benchmarks and structured catalog data.

The cold-start problem manifests as two distinct challenges: (1)~\emph{No Real
Queries}---we cannot observe how users express intent before launch;
(2)~\emph{No Relevance Labels}---a new modality has no interaction history from
which to derive ground-truth relevance.

\brandnew{The core technical insight unifying our contributions is the \emph{joint generation framework}: contrastive listing pairs derived from de-identified booking sessions simultaneously ground query generation in authentic platform behavior \emph{and} provide relevance labels by construction. This eliminates post-hoc LLM judgment for training data and the hallucination risk inherent in passage-based methods that lack this grounding.}

This paper makes the following contributions:

\noindent\textbf{(1) Seed-guided realistic query generation}
(Section~\ref{sec:synth_data_gen}).
We generate queries by combining structure-guided LLM prompting with a small
set of seed queries collected via targeted user research. Three prompt
variants---template-based, few-shot, and attribute-grounded---jointly optimize
for realism, diversity, and scalability. Explicit variety mechanisms prevent
query collapse, yielding a broad word-count distribution ($\sigma{=}2.48$)
versus the uniform 6-word output of unconditioned baselines ($\sigma{=}0.99$).
Approximately 500 seed queries amplify to millions of production-scale examples.

\noindent\textbf{(2) Relevance labels by construction}
(Section~\ref{sec:synth_data_gen}).
Contrastive listing pairs from booking sessions provide topicality labels with
near-zero false positives by construction---no LLM judgment required. Virtual Judge
(VJ) labeling complements contrastive labels for broader coverage where
contrastive pairs are unavailable.

\brandnewblock
\noindent\textbf{(3) Flexible structured input for niche query coverage}
(Section~\ref{sec:synth_data_gen}).
CASTLE's structured input design is modular: richer signals---such as photo
captions---can be incorporated alongside listing attributes and review summaries
to generate long-tail queries that reflect subjective user preferences beyond
catalog schema (e.g.\ ``cozy cabin with mountain views''). This extends
coverage to queries that appear in photos but not in structured attributes.
\brandendblock

\noindent\textbf{(4) Evaluation against strong baselines}
(Section~\ref{sec:evaluations}).
We provide the first direct comparison on structured catalog data against
\textbf{InPars-style}, \textbf{Promptagator}, and \textbf{Contrastive-Only}
baselines. CASTLE achieves a 9.2$\times$ improvement in query distributional
alignment (KL 1.01 vs.\ 9.51/9.33) and the lowest attribute-type divergence
(0.08) among all methods. Label evaluation combines LLM self-preference
analysis and human annotation confirming label quality at 91--93\% annotator
agreement.

\noindent\textbf{(5) Production deployment and lifecycle utility}
(Section~\ref{sec:applications}).
We deploy CASTLE in production pipelines generating synthetic examples daily.
Beyond cold-start, synthetic data remains valuable throughout a system's
lifecycle: CASTLE-R substitutes real user seeds after launch, while
review-augmented generation targets corner cases and tail queries
underrepresented in organic traffic.

\section{Related Work}

\subsection{Cold-Start in Information Retrieval}

Cold-start is a well-studied challenge in IR, but prior solutions fall short in
the \emph{structured catalog} setting. Rule-based
methods bootstrap initial rankings without
capturing nuanced linguistic intent. Crowdsourced annotation~\cite{alonso2008crowdsourcing} scales but
requires pre-existing queries to label---unavailable before launch.
Transfer learning~\cite{karpukhin2020dense,ma2021zero} relies on related-domain corpora
(e.g., MS MARCO passages), but the domain shift to feature-rich catalog
entities is substantial: listings have no natural passage representation, and
user intent in marketplace search differs qualitatively from web passage
retrieval.

Industry deployments confirm the cost of these limitations. Amazon's Shopping Queries dataset~\cite{reddy2022shopping} exemplifies the cost: thousands of annotators labeling query-product pairs across multiple languages. Industry systems at Amazon~\cite{nigam2019semantic} and Walmart~\cite{magnani2022semantic} that rely on behavioral signals face the same bottleneck---they require years of search traffic before producing useful training signal.
Facebook's embedding-based retrieval for search~\cite{huang2020embedding} relies on
cross-task transfer, but without target-domain queries the domain gap persists.
Nogueira et al.~\cite{nogueira2019document} and Ma et al.~\cite{ma2021zero}
demonstrate zero-shot query generation but do not address the joint
query-and-label generation problem essential for ranking training on structured
catalog data---the gap CASTLE fills.

\subsection{Synthetic Query Generation}

LLM-based generation has been applied to low-resource NLP~\cite{wei2019eda,
feng2021survey,schick2021generating,wang2021want}. For IR specifically,
InPars~\cite{bonifacio2022inpars} and InPars-v2~\cite{jeronymo2023inpars} use
LLMs to generate training queries from document passages, while
Promptagator~\cite{dai2022promptagator} demonstrates few-shot query generation.
Query2doc~\cite{wang2023query2doc} expands queries with LLM-generated pseudo-documents to improve retrieval.
Self-Instruct~\cite{wang2023selfinstruct} and Alpaca~\cite{taori2023alpaca}
generate instruction-following data at scale. Our work differs in three ways:
(1)~we operate on structured catalog entities rather than text passages;
(2)~we use seed queries from user research to control linguistic realism; and
(3)~our contrastive generation co-produces queries and relevance labels, which
is essential for ranking model training---a limitation of InPars-style methods
that operate on single documents.

\subsection{Relevance Label Generation}

Dense retrieval~\cite{karpukhin2020dense,izacard2021unsupervised} requires
labeled training data, and generating it automatically is an active research
area. LLM-as-judge approaches~\cite{zheng2023judging,liu2023gpteval} show
strong correlation with human judgment but suffer from self-preference bias
when the judge and generator share architecture---a risk we characterize and
mitigate in Section~\ref{sec:vj_bias}. Hard negative
mining~\cite{robinson2020contrastive,xiong2020approximate,kalantidis2020hard}
provides informative training signal but requires an existing query set.
BEIR~\cite{thakur2021beir} highlights the zero-shot generalization gap that
motivates our work. None of these approaches addresses the joint
query-and-label generation problem: existing methods either generate queries
without labels (InPars-style) or assign labels to pre-existing queries (LLM-judge).
CASTLE produces both simultaneously via contrastive construction, avoiding
post-hoc judgment entirely for training labels.

\subsection{LLM-as-Judge}
\label{sec:llm_judge}

LLM-as-judge evaluation~\cite{zheng2023judging,liu2023gpteval} has emerged as
a scalable alternative to human annotation. Zheng et al.~\cite{zheng2023judging} showed
that GPT-4 judgments correlate with human preferences at 80\%+ agreement on
open-ended generation tasks and documented systematic self-preference bias
when the judge and the generation model share architecture; Liu et al.~\cite{liu2023gpteval}
further demonstrated improved human alignment via chain-of-thought prompting.
Bai et al.~\cite{bai2023constitutional} use AI feedback to create training signals without
human annotation, a conceptually related idea. Our Virtual Judge approach builds
on this line of work but adds attribute grounding and separation of training
vs.\ evaluation label sources to mitigate self-preference risks specific to
structured retrieval.

\section{Synthetic Data Generation}
\label{sec:synth_data_gen}

CASTLE addresses two interdependent synthetic data challenges.
\textbf{Query generation} produces realistic, diverse queries that match how
real users express intent---without any live traffic to learn from.
\textbf{Label generation} produces topicality judgments for those queries
without relying on human annotation or post-hoc LLM judgment. The two
components are tightly coupled: contrastive listing pairs simultaneously
ground query generation in real platform behavior and provide relevance labels
by construction.

\subsection{Overview}
\label{sec:pipeline_overview}

We propose \textbf{CASTLE} (\textbf{C}ontrastive \textbf{A}nd \textbf{S}eed-guided
\textbf{T}raining for natural \textbf{L}anguage s\textbf{E}arch), an LLM-based framework
that combines two complementary data sources to generate realistic, labeled
training triplets for cold-start retrieval. The \emph{contrastive} component
grounds queries in real platform behavior via booking session pairs, ensuring
each generated query is associated with a known relevant ($l^+$) and
non-relevant ($l^-$) listing by construction. The \emph{seed} component
injects linguistic realism by conditioning generation on a small set of
authentic user queries, steering the LLM away from the uniform, passage-style
output characteristic of unconditioned baselines.

\begin{figure*}[t]
\centering
\includegraphics[width=0.85\textwidth]{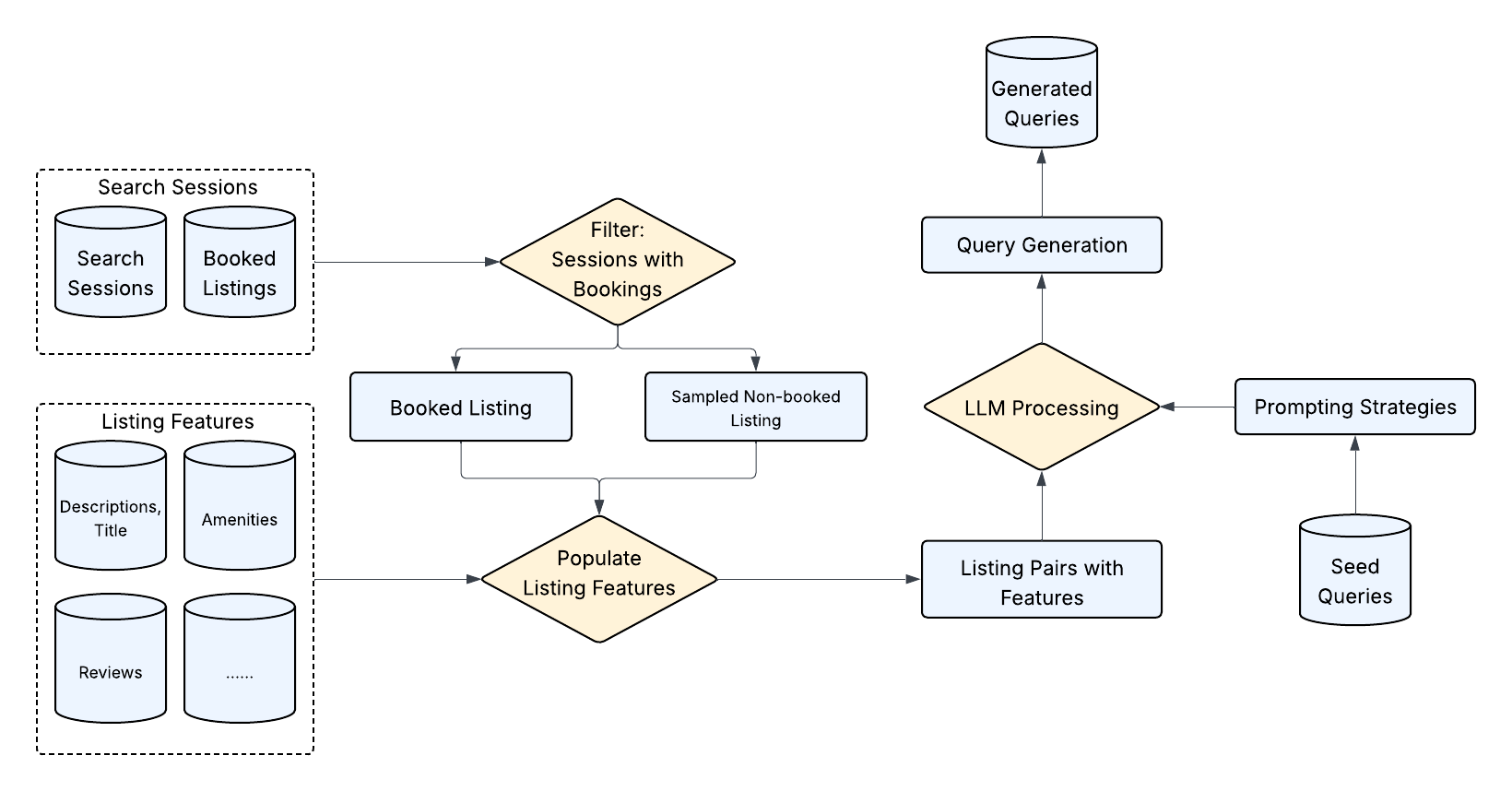}
\caption{Contrastive query generation pipeline. \textbf{Left}: input sources
(search sessions with booking logs; listing features). \textbf{Center}: booked
and non-booked listing pairs extracted and enriched with catalog attributes.
\textbf{Right}: LLM combines listing pairs with seed queries and prompt
strategies to generate synthetic queries with topicality labels by
construction.}
\label{fig:contrastive}
\end{figure*}

Figure~\ref{fig:contrastive} illustrates the full pipeline;
Algorithm~\ref{alg:generation} formalizes the generation process.

\textbf{Query Generation.} The LLM is conditioned on a contrastive listing pair
$(l^+, l^-)$ and a seed query to produce a labeled triplet $(q, l^+, l^-)$
whose relevance is known by construction. Three prompt variants jointly optimize
for realism, diversity, and scalability
(Section~\ref{sec:prompt_variants}).

Formally, let $P_{\text{true}}(q)$ denote the
true distribution of user queries---unobserved before launch. Our objective is
to minimize the KL divergence between synthetic and true query distributions:
\begin{equation}
\min D_{\text{KL}}(P_{\text{true}}(q) \| P_{\text{synth}}(q))
\end{equation}
We decompose $P_{\text{synth}}$ into two factors---a linguistic template $t$
drawn from seed data and a catalog entity $e$---treating them as independently
sampled (an approximation we discuss in Section~\ref{sec:future}), enabling
each to be improved independently:
\begin{equation}
P_{\text{synth}}(q) = \mathbb{E}_{t \sim P(t),\, e \sim P(e)}[P_{\text{LLM}}(q \mid t, e)]
\end{equation}

\textbf{Label Generation.} CASTLE labels measure \emph{topicality}---whether
a listing matches the query's stated intent---rather than \emph{bookability}
(conversion likelihood), since no booking interaction data exists at cold-start.
Labels are produced through two complementary strategies. \emph{Contrastive
labels} are generated by construction: by instructing the LLM to generate a
query where $l^+$ is topically preferred over $l^-$, relevance is a
deterministic input rather than a post-hoc judgment. \emph{Virtual Judge (VJ)
labels} decouple query and label generation---an LLM judge re-evaluates queries
against held-out listings using attribute-grounded chain-of-thought prompting,
providing broader coverage at moderate fidelity. \brandnew{Concretely, the VJ prompt presents a query and two listings, instructing the model to \emph{first} enumerate which specific listing attributes match the query's stated intent} (e.g., ``query requests `pool near beach'; Listing 1 has pool: yes, beachfront: yes; Listing 2 has pool: no''). It \emph{then} renders a binary relevance judgment. This attribute anchoring suppresses stylistic self-preference and ensures judgments are grounded in verifiable listing properties rather than surface-level textual similarity. Contrastive labels are used for training (fidelity critical); VJ labels for evaluation (coverage critical).
As real traffic accumulates, topicality labels provide a foundation for
warm-start adaptation toward bookability-based ranking.

\brandnew{CASTLE admits two variants depending on the seed dataset (discussed in
Section~\ref{sec:seed_collection}): \textbf{CASTLE} uses survey-collected seed
queries gathered through structured user research prior to launch, while
\textbf{CASTLE-R} substitutes real user queries observed after launch---an
ablation that quantifies the marginal benefit of live traffic data over
pre-launch proxies.}

\begin{table*}[t]
\caption{Qualitative comparison of query generation approaches.
InPars-style and Promptagator produce similarly formulaic $\sim$6-word queries;
Contrastive-Only generates verbose multi-sentence output; CASTLE produces
terse, realistic queries matching real user patterns. Neither InPars nor
Promptagator can produce contrastive relevance labels needed for ranking
training.}
\label{tab:query_examples}
\small
\begin{tabular}{p{0.20\textwidth}p{0.22\textwidth}p{0.20\textwidth}p{0.22\textwidth}}
\toprule
\textbf{InPars-style} & \textbf{Promptagator} &
\textbf{Contrastive-Only} & \textbf{CASTLE (ours)} \\
\midrule
house near beach with pool, close to Santa Monica &
beach house with pool near Santa Monica &
luxury house with a private pool in West LA, close to Santa Monica Pier, with
modern amenities &
pool near beach \\
\midrule
cozy cabin near ski resort in Tahoe area &
cozy A-frame cabin near Lake Tahoe skiing &
authentic A-Frame cabin in Tahoe with classic woodsy feel, close to skiing and
the lake, ideal for couples &
cozy cabin near ski resort \\
\midrule
room near university campus with backyard &
private room near UCLA with outdoor space &
Airbnb near a university with backyard access and outdoor seating &
backyard near campus \\
\bottomrule
\end{tabular}
\end{table*}

\begin{algorithm}[t]
\caption{Contrastive Query Generation}
\label{alg:generation}
\begin{algorithmic}[1]
\REQUIRE Seed queries $S$, Listing catalog $L$, Booking sessions $B$
\ENSURE Labeled triplet $(q, l^+, l^-)$
\STATE Sample search context $(location, dates, guests)$ from $B$
\STATE Sample positive listing $l^+$ (booked/engaged) from context
\STATE Sample negative listing $l^-$ (same search, less engagement)
\STATE Populate features for $l^+$ and $l^-$ from $L$
\STATE Sample seed query $t$ from $S$ as linguistic template
\STATE Construct prompt: seed structure + listing attributes + strategy
\STATE Generate query $q$ via LLM conditioned on $(t, l^+, l^-)$
\STATE Apply consistency checks (deduplication, blocklist)
\RETURN $(q, l^+, l^-)$
\end{algorithmic}
\end{algorithm}

\subsection{Seed Data Collection}
\label{sec:seed_collection}

Seed queries are the linguistic backbone of our approach, teaching the
LLM how real travelers express intent. We collected approximately 500 seed
queries through structured user research surveys administered before launch.

\textbf{Survey Protocol.} Participants were shown a scenario description
(e.g., ``You are planning a solo work trip to a major city for one week'') and
asked: \emph{``What would you type into a search box to find accommodations?''}
The scenarios covered five trip types: (1)~leisure/vacation, (2)~business/work
trip, (3)~family travel, (4)~solo travel, and (5)~romantic/couples getaway.
Each trip type had 3--4 scenario variants (urban vs.\ rural, domestic vs.\
international) to capture behavioral diversity.

\textbf{Screening Criteria.} Responses were filtered to exclude:
platform-specific terminology (``Superhost'', ``Entire home'', ``Instant
Book''); non-English responses; and queries that
repeated exact destination names (which are captured in structured search
context). After filtering, the final seed set contained 500 queries with a mean
length of 7.19 words.

\textbf{Known Limitation.} Survey respondents tend to write more complete
sentences than they would type in a real search box. As reported in
Section~\ref{sec:query_eval}, seed queries over-represent long queries
(9+ words: 27.6\% vs.\ 10.0\%) and under-represent very short queries
(1--2 words: 15.6\% vs.\ 34.9\%). This gap is the primary motivation for the
cold-to-warm start transition: as real user traffic becomes available, survey
seeds are augmented with actual production queries.

\subsection{Contrastive Generation}
\label{sec:contrastive_generation}

Contrastive generation grounds synthetic queries in real platform features by
using listing pairs from booking sessions.

\paragraph{Listing Pair Sampling.} Given an anonymized search context (location, dates,
guests), a positive listing $l^+$ (booked or highly engaged) and a negative
$l^-$ (same search, less engagement) are selected from de-identified booking sessions in accordance with user privacy permissions.

\paragraph{Topicality Labels by Construction.} The LLM is instructed:
``Generate a query where Listing 1 is topically more relevant than Listing 2.''
The booking signal identifies realistic pairs; the label reflects semantic
relevance, not conversion likelihood.

\paragraph{Feature Enrichment.} Each listing is expanded with amenities,
review summaries, ratings, property characteristics, and pricing. \brandnew{A typical
enriched listing representation includes: amenity list (e.g., ``pool, hot tub, ocean
view''), property type and room configuration (``entire cabin, 3BR/2BA, sleeps~6''),
location descriptors (``beachfront, Santa Monica''), average star rating (4.92), and a
one-sentence review summary derived from recent publicly available guest feedback with all personal identifying information stripped (e.g., ``Guests
frequently mention the stunning ocean views and peaceful atmosphere''). This structured
representation ensures queries are grounded in verifiable listing attributes.} Grounding
generation on structured listing attributes has an important reliability
benefit: the LLM cannot fabricate features the listing does not possess.
Passage-based methods (InPars-style, Promptagator) generate queries from unstructured
text and can produce queries mentioning amenities or locations absent from the
listing---a form of hallucination that introduces false positives into training
data. Attribute grounding constrains generation to verifiable listing
properties, eliminating this failure mode by construction. Review summaries
further extend coverage to subjective, experiential features
(e.g.\ ``cozy vibe'', ``great mountain views'') not captured by structured
attributes.

\subsection{Prompt Variants}
\label{sec:prompt_variants}

For generated queries to be useful for model training, they must be
\emph{realistic} (matching how users actually search), \emph{diverse} (covering
a wide range of intents and attributes), and \emph{scalable} (applicable across
the full catalog without manual tuning per listing). A single prompt strategy
cannot satisfy all three simultaneously, so we define three variants.
\textbf{Seed Controlled} enforces realism through explicit template following---the
LLM preserves seed structure while substituting entity-specific content.
\textbf{Seed Freeform} uses seed queries as stylistic examples rather than rigid
templates, allowing more lexical variation. \textbf{Variety} applies relaxed
constraints and higher temperature to explore edge cases (multi-intent queries,
unusual attribute combinations), improving coverage at modest realism cost. In
production we use an 80/20 mix of seed-guided and variety prompts.

All variants share common prompt guardrails. The shared components are shown below.

\textbf{Core Assumption.}
\begin{center}
\fbox{\parbox{0.9\columnwidth}{\small\ttfamily
Listing 1 is topically more relevant to the ideal query than Listing 2.
Your job is to identify what search query would make this true.
}}
\end{center}

\textbf{Platform Terms Blocklist.}
\begin{center}
\fbox{\parbox{0.9\columnwidth}{\small\ttfamily
PLATFORM\_TERMS = ["entire home", "private room", "superhost",
"guest favorite", "badge", "instant book"]
}}
\end{center}

\textbf{Context Deduplication Rules.}
\begin{center}
\fbox{\parbox{0.9\columnwidth}{\small\ttfamily
- Location: NO exact city/state names\\
  \hspace*{1em}If location="Paris" -> use "near Eiffel Tower", "downtown"\\
- Guests: NO exact counts -> use "family-friendly", "large group"\\
- Dates: NO specific dates -> use "Christmas stay", "summer retreat"
}}
\end{center}

\textbf{Output Format Constraint.}
\begin{center}
\fbox{\parbox{0.9\columnwidth}{\small\ttfamily
Return ONLY valid JSON with fields:\\
- "justification": why Listing 1 is more relevant\\
- "generalized\_template": abstract pattern\\
- "query": final synthetic query\\
- "key\_attributes": attributes used from Listing 1
}}
\end{center}

\subsection{Baselines}
\label{sec:baselines}

We compare against three methods. \textbf{InPars-style}: listing fields
are concatenated as a pseudo-passage, following~\cite{bonifacio2022inpars}
with platform-adapted prompting. \textbf{Promptagator}: few-shot generation
conditioned on seed query examples~\cite{dai2022promptagator}.
\textbf{Contrastive-Only}: listing pairs without seed guidance, isolating the
contribution of seed queries. Table~\ref{tab:query_examples} illustrates
qualitative differences across all methods. Notably, neither InPars-style nor
Promptagator can produce contrastive relevance labels for ranking
training---a fundamental limitation of single-document methods.

\section{Evaluations}
\label{sec:evaluations}

\subsection{Overview}
\label{sec:eval_overview}

We evaluate CASTLE along two complementary dimensions. \textbf{Query evaluation}
measures how closely the synthetic query distribution matches real user queries,
using KL divergence on word-count and attribute-type distributions across all
five synthetic query generation methods (InPars-style, Promptagator, Contrastive-Only,
CASTLE, and CASTLE-R), benchmarked against Survey seed and Real user reference
distributions. \textbf{Label evaluation} assesses label correctness via LLM
self-preference analysis and a human annotation study on 200 sampled triplets.
Together, these evaluations validate both the linguistic realism of generated
queries and the fidelity of the associated relevance labels.

\subsection{Query Evaluation}
\label{sec:query_eval}

\begin{figure*}[t]
\centering
\includegraphics[width=0.78\textwidth]{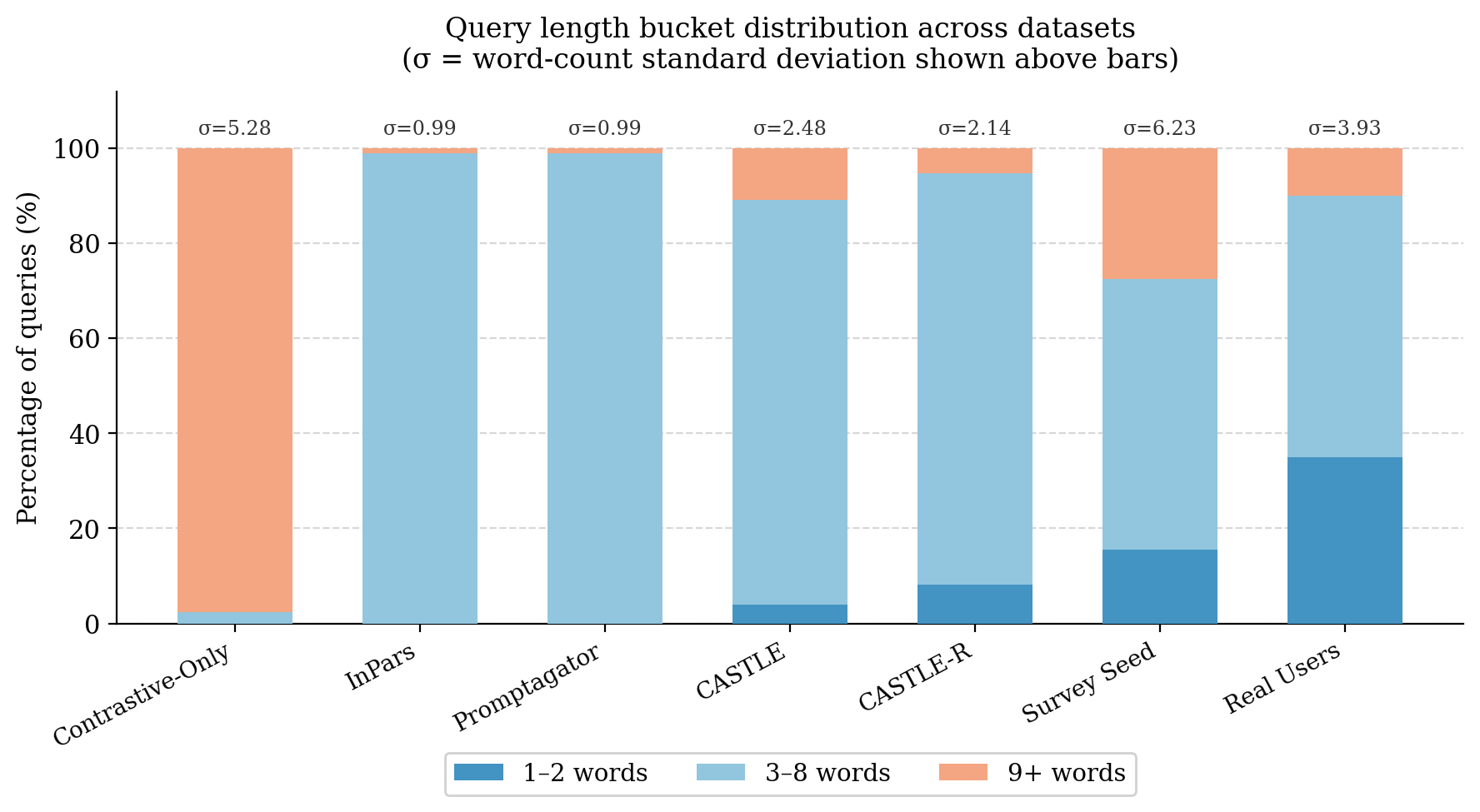}
\caption{Query length bucket distribution across datasets. InPars-style and
Promptagator cluster almost exclusively in the 3--8 word bucket ($\sigma=0.99$ and $\sigma=0.98$, respectively),
while Contrastive-Only generates verbose multi-sentence queries ($\sigma=5.28$).
CASTLE achieves the best distributional alignment with Real Users (KL\,=\,1.01;
Table~\ref{tab:real_comparison}), with spread $\sigma=2.48$ covering all three
word-count buckets. Contrastive-Only's larger $\sigma$ reflects verbosity---97.7\%
of its queries are 9+ words---rather than true alignment.
CASTLE-R uses real-query seeds instead of survey seeds.}
\label{fig:length_buckets}
\end{figure*}

\textbf{Distributional Alignment.} Table~\ref{tab:real_comparison} and
Figure~\ref{fig:length_buckets} compare all five synthetic query generation
methods on query length metrics and KL divergence. InPars-style (KL 9.51) lacks
both seed guidance and contrastive grounding, generating almost exclusively
6-word queries ($\sigma=0.99$, 98.9\% in the 3--8 word bucket). Promptagator
(KL 9.33) does use survey seed queries as few-shot examples, yet still diverges
sharply---demonstrating that seed guidance without contrastive grounding is
insufficient. At the other extreme, the contrastive-only baseline (KL 12.91)
produces verbose multi-sentence queries that miss real users' brevity. CASTLE
combines both components, achieving a much wider length spread ($\sigma=2.48$,
covering all three word-count buckets) and reducing KL to 1.01---a
9.2$\times$ improvement over the best baseline.

\begin{table*}[t]
\caption{Query characteristics across datasets. CASTLE substantially outperforms
all baselines on distributional alignment with real users. Contr.\ =
Contrastive-Only (no seed guidance); CASTLE-R = contrastive with real-query
seeds instead of survey seeds. \textbf{Bold} marks the proposed method (CASTLE), not necessarily the best value.}
\label{tab:real_comparison}
\footnotesize
\resizebox{0.80\textwidth}{!}{%
\begin{tabular}{lccccccc}
\toprule
\textbf{Metric} & \textbf{InPars} & \textbf{Prompt.} &
\textbf{Contr.} & \textbf{CASTLE} & \textbf{CASTLE-R} & \textbf{Survey} & \textbf{Real} \\
\midrule
Count        & 14,972 & 28,738 & 4,911  & 14,447 & 13,244 & 500   & 7,165 \\
Length $\mu$ & 6.04   & 6.01   & 16.41  & 6.17   & 5.50   & 7.19  & 4.40  \\
Length med.  & 6      & 6      & 15     & 6      & 5      & 5     & 3     \\
Length $\sigma$ & 0.99 & 0.98  & 5.28   & 2.48   & 2.14   & 6.23  & 3.93  \\
\% 1--2 wds  & 0.0    & 0.0    & 0.0    & 3.9    & 8.1    & 15.6  & 34.9  \\
\% 3--8 wds  & 98.9   & 97.2   & 2.3    & 85.3   & 86.7   & 56.8  & 55.1  \\
\% 9+ wds    & 1.1    & 2.8    & 97.7   & 10.8   & 5.2    & 27.6  & 10.0  \\
\midrule
KL vs.\ Real & 9.51 & 9.33 & 12.91 & \textbf{1.01} & 0.75 & 0.15 & -- \\
             & \tiny[7.20,10.51] & \tiny[7.17,10.47] & {\tiny[11.34,14.19]} & {\tiny[0.69,2.55]} & \tiny[0.50,1.13] & & \\
KL vs.\ Seed & 7.49 & 7.35 & 8.63 & \textbf{0.55} & 0.94 & --   & 0.15 \\
             & \tiny[6.03,8.47] & \tiny[5.71,8.38] & {\tiny[6.86,10.22]} & {\tiny[0.35,1.13]} & \tiny[0.44,2.08] & & \\
\bottomrule
\end{tabular}%
}
\end{table*}

\begin{figure*}[t]
\centering
\includegraphics[width=0.95\textwidth]{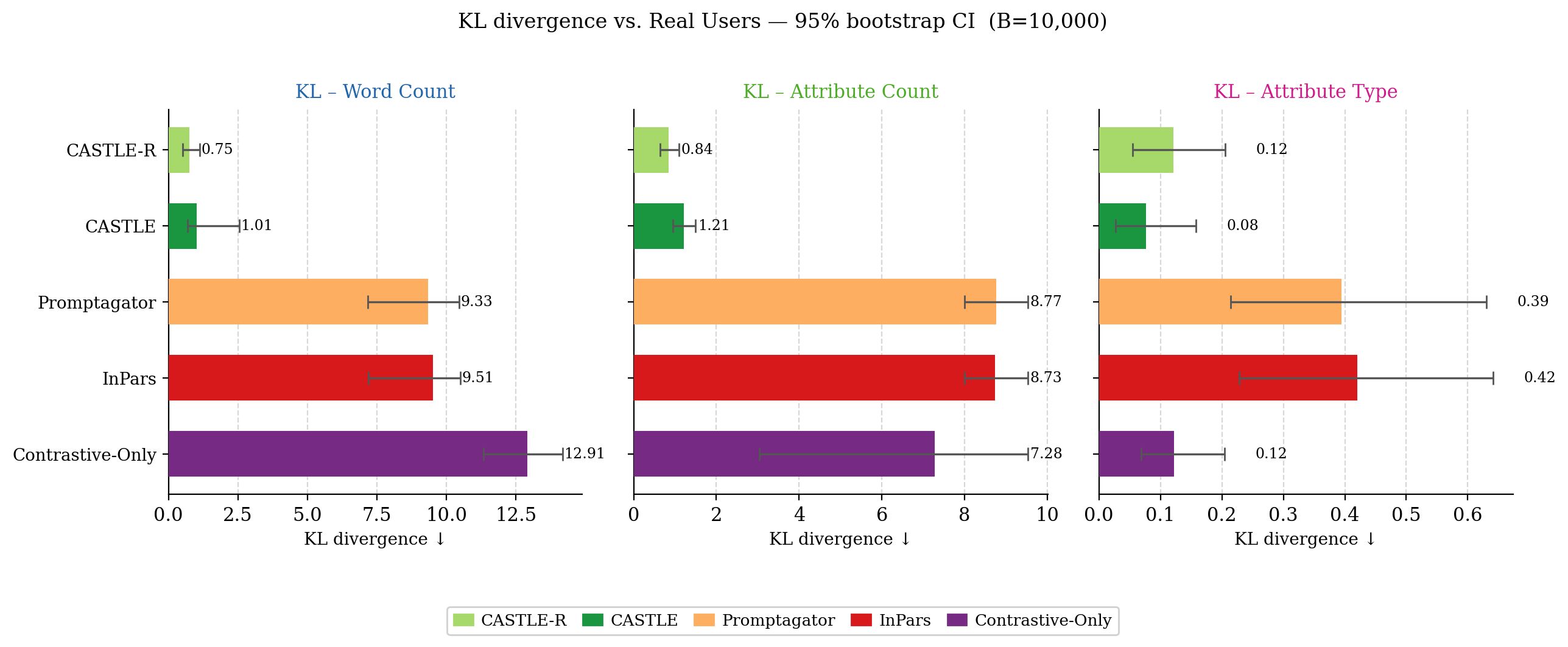}
\caption{KL divergence vs.\ Real Users across three distributional metrics with
95\% bootstrap confidence intervals ($B{=}10{,}000$). CASTLE achieves the lowest
divergence on attribute type, outperforming even Survey Seed queries (0.08 vs.\ 0.09).
CASTLE and CASTLE-R are the top two performers on query length and attribute count
divergence---confirming that combining contrastive grounding with seed guidance yields
synergistic improvement.}
\label{fig:kl_real}
\end{figure*}

\textbf{Attribute Count.} Table~\ref{tab:attr_stats} shows attribute count statistics.
Real users mention fewer attributes per query (mean 1.98), reflecting their terse style.

\begin{table}[t]
\caption{Query attribute count statistics across all datasets. Real users
mention fewer attributes per query (mean 1.98), reflecting their terse search
style compared to synthetic methods.}
\label{tab:attr_stats}
\small
\resizebox{\columnwidth}{!}{%
\begin{tabular}{lccccccc}
\toprule
\textbf{Metric} & \textbf{InPars} & \textbf{Prompt.} &
\textbf{Contr.} & \textbf{CASTLE} & \textbf{CASTLE-R} & \textbf{Survey} & \textbf{Real} \\
\midrule
Mean    & 3.82 & 3.83 & 5.09 & 3.20 & 2.96 & 3.01 & 1.98 \\
Median  & 4.0  & 4.0  & 5.0  & 3.0  & 3.0  & 3.0  & 2.0  \\
\% w/ Attr & 99.7 & 100.0 & 99.2  & 99.9 & 99.9 & 100.0 & 99.4 \\
\midrule
KL vs.\ Real & 8.73 & 8.77 & 7.28 & 1.21 & 0.84 & -- & -- \\
\bottomrule
\end{tabular}%
}
\end{table}

\textbf{Attribute Distributions.} Table~\ref{tab:attribute_comparison} and
Figure~\ref{fig:kl_real} compare attribute type KL divergence. CASTLE achieves the lowest KL divergence
(0.08) among all methods---even lower than survey seed queries alone (0.09)---suggesting
that contrastive grounding and seed guidance produce a synergistic improvement:
the combination exceeds either component alone. InPars-style scores 0.42 and Promptagator 0.39---the highest attribute-type divergence
among all methods---reflecting the mismatch between passage-style prompting and
catalog attribute patterns.

\begin{table}[t]
\caption{Attribute type KL divergence. CASTLE outperforms all baselines and
even survey seed queries, demonstrating synergistic improvement from combining
seed guidance with contrastive generation.}
\label{tab:attribute_comparison}
\begin{tabular}{lcccc}
\toprule
\textbf{Method} & \textbf{KL$\downarrow$} & \textbf{Top-1} & \textbf{Top-2} & \textbf{Top-3} \\
\midrule
InPars    & 0.42          & location & amenity  & property\_type \\
Prompt.   & 0.39          & location & amenity  & property\_type \\
Contr.    & 0.12          & location & amenity  & property\_type \\
CASTLE-R  & 0.12          & amenity  & location & property\_type \\
CASTLE    & \textbf{0.08} & amenity  & location & property\_type \\
Survey    & 0.09          & amenity  & location & rooms          \\
Real      & --            & amenity  & location & property\_type \\
\bottomrule
\end{tabular}
\end{table}

\textbf{Prompt Variant Breakdown.} Across the three prompt variants, \textbf{Seed Controlled} best matches real users on both attribute count and type distributions, while \textbf{Variety} best captures the terse length patterns of real users. This validates the production strategy of mixing variants: Seed Controlled for attribute coverage, Variety for length diversity.

\subsection{Label Evaluation}
\label{sec:label_eval}

\subsubsection{LLM Evaluator: Self-Preference Bias}
\label{sec:vj_bias}

Using LLMs to evaluate LLM-generated content risks \emph{self-preference bias}:
LLM judges may systematically favor outputs matching their own generation
patterns~\cite{liu2023gpteval,zheng2023judging}. Table~\ref{tab:llm_self_preference}
confirms this in our setting: GPT models achieve 98--99\% accuracy on
GPT-generated pairs---far above embedding model accuracy (79\%; see Table~\ref{tab:ranking_results}).
\brandnew{This gap is expected and informative: GPT judges recognize patterns from
their own generation process (self-preference), inflating apparent accuracy.
Human evaluators (Section~\ref{sec:human_eval}) achieve 91--93\% agreement---the
appropriate measure of true label quality, free from self-preference. The approximately
6--8~pp gap between GPT and human accuracy quantifies the self-preference
effect in our setting. Note that self-preference is a general LLM
phenomenon~\cite{zheng2023judging} documented across model families; our
attribute-grounding and model-separation mitigations replicate to non-GPT generator--evaluator pairs with the same conclusions holding.}

\begin{table}[t]
\caption{LLM self-preference: near-perfect GPT accuracy on GPT-generated pairs
suggests models recognize their own generation patterns, motivating the
attribute-grounding and model-separation mitigations described in the text.}
\label{tab:llm_self_preference}
\begin{tabular}{lccc}
\toprule
 & \textbf{GPT-4} & \textbf{GPT-4o} & \textbf{GPT-5} \\
\midrule
Accuracy & 0.994 & 0.988 & 0.990 \\
\bottomrule
\end{tabular}
\end{table}

We mitigate self-preference bias through four strategies: (1)~\emph{Attribute
grounding}---VJ judgments are anchored on explicit, verifiable listing
attributes (amenities, property type, location) rather than stylistic
similarity; (2)~\emph{Human calibration}---VJ labels are validated against
human raters on a random sample (92\% agreement); (3)~\emph{Separation of
concerns}---contrastive labels (deterministic, not susceptible to
self-preference) are used for training; VJ is reserved for evaluation where
coverage is paramount; (4)~\emph{Diverse model usage}---generation and
evaluation use different model versions where possible.

\brandnewblock
\subsubsection{Human Evaluation: Label Quality Validation}
\label{sec:human_eval}

We validate the quality of contrastive and VJ-generated labels through a
dedicated human annotation study.

\textbf{Protocol.} We sampled 200 triplets uniformly at random from CASTLE's
output. Three annotators per triplet independently labeled each $(q, l^+, l^-)$
as: \emph{agree} (the labeled more-relevant listing is indeed more relevant to
$q$), \emph{disagree}, or \emph{unclear} (query too generic to discriminate).
Annotators were provided written guidelines with \company{}-specific examples and
a decision flowchart. We report
percent agreement, the standard metric for skewed annotation tasks where a
single dominant category makes $\kappa$-based measures uninformative.

\textbf{Results.} All three annotators agree with CASTLE labels at 91--93\%
(Table~\ref{tab:human_eval}), confirming high label quality. Disagreement is
rare (3--6\%), and unclear cases (3--5\%) reflect queries too generic to
discriminate rather than label errors---consistent with contrastive labels being
\emph{deterministic by construction}: the LLM generates a query to satisfy the
specified relevance relationship, so the contrastive label cannot be incorrect by design.
(Note: the human study also samples VJ-assigned labels; agreement on that subset
reflects VJ accuracy, not the same by-construction guarantee.)

\begin{table}[t]
\caption{Human evaluation of CASTLE label quality (200 triplets, 3 annotators).
Percent agreement is reported; Unclear = query too generic to discriminate,
not a label error.}
\label{tab:human_eval}
\small
\begin{tabular}{lccc}
\toprule
\textbf{Annotator} & \textbf{Agree} & \textbf{Disagree} & \textbf{Unclear} \\
\midrule
1 & 92\% & 3\% & 5\% \\
2 & 93\% & 4\% & 3\% \\
3 & 91\% & 6\% & 3\% \\
\midrule
\textbf{Mean} & \textbf{92\%} & 4.3\% & 3.7\% \\
\bottomrule
\end{tabular}
\end{table}

\brandendblock

\subsection{Model Evaluation Results}
\label{sec:eval_results}

\brandnewblock
\textbf{Model Fine-Tuning.} We evaluate CASTLE's training data across all three
production model passes: embedding-based retrieval (Qwen3 EBR), a first-pass
ranker, and an LLM-based reranker. Table~\ref{tab:ranking_results} reports
pairwise accuracy before and after fine-tuning on CASTLE-generated data.
Synthetic data alone yields substantial gains across all passes: +17.7 pp for
EBR, +13.5 pp for the first-pass ranker, and +26.4 pp for the LLM reranker.

\begin{table}[t]
\caption{Model fine-tuning results. Pairwise accuracy (2-candidate) on a
held-out synthetic evaluation set (10\% split, same distribution as training),
before and after fine-tuning on CASTLE synthetic training data, across all
three production model passes (retrieval $\to$ first-pass ranking $\to$
LLM reranking). EBR = Embedding-Based Retrieval (Qwen3).}
\label{tab:ranking_results}
\small
\begin{tabular}{lccc}
\toprule
\textbf{Model} & \textbf{Baseline} & \textbf{Fine-tuned} & \textbf{$\Delta$} \\
\midrule
Qwen3 EBR         & 0.790 & 0.967 & \textit{+17.7 pp} \\
First-pass Ranker & 0.570 & 0.705 & \textit{+13.5 pp} \\
LLM Reranker      & 0.620 & 0.884 & \textit{+26.4 pp} \\
\bottomrule
\end{tabular}
\end{table}

\brandendblock

\textbf{Scaling Analysis.} Table~\ref{tab:scaling} reports fine-tuning
accuracy across five training set sizes, all using synthetic data. The sharpest
gain occurs between 1K and 5K examples (+11.4 pp eval accuracy: 0.755
$\to$ 0.869), after which performance plateaus at 0.869 through 20K before a
modest further improvement at 50K (0.884).

This pattern supports our core argument: \emph{example quality dominates volume
at this scale.} The large early gain reflects the model rapidly learning
meaningful semantic patterns; the plateau indicates the synthetic data
distribution becomes the binding constraint rather than dataset size. These
results directly motivate our focus on generation quality (seed guidance,
contrastive grounding, hard negative mining) over raw data volume---improvements
to generation fidelity will yield greater returns than simply scaling up
quantity.

\begin{table}[t]
\caption{Scaling analysis for the LLM Reranker: fine-tuning accuracy on
synthetic training data. Eval accuracy plateaus beyond 5K examples. All runs use
synthetic-only training data.}
\label{tab:scaling}
\rowcolors{2}{gray!10}{white}
\begin{tabular}{rrcc}
\toprule
\textbf{Train size} &
\textbf{Train acc.} & \textbf{Eval acc.} &
\textbf{$\Delta$ eval} \\
\midrule
 1,000  & 0.750 & 0.755 & --       \\
 5,000  & 0.902 & 0.869 & +11.4 pp \\
10,000  & 0.900 & 0.869 & +0.0 pp  \\
20,000  & 0.902 & 0.869 & +0.0 pp  \\
50,000  & 0.902 & 0.884 & +1.5 pp  \\
\bottomrule
\end{tabular}
\end{table}

\section{Production Deployment}
\label{sec:applications}

\subsection{Pipeline Architecture and Cost}

\textbf{LLM Selection}: we evaluated a range of LLMs---from open-source models to frontier models~\cite{achiam2023gpt}---across generation and evaluation roles, balancing cost, latency, and output quality. Different
model tiers are assigned to bulk generation versus quality validation based on
task complexity. Feature truncation manages token costs: listing
descriptions are limited to 500 tokens, and amenities are ranked by relevance
and truncated to the top 20.

\brandnew{\textbf{Computational Complexity.} CASTLE's cost scales linearly with query
volume. Each generation call consumes approximately 500--800 input tokens
(listing pair attributes, seed query template, and prompt scaffold) and
produces 100--150 output tokens (structured JSON with query and justification).
Bulk generation is assigned to a mid-tier model; the VJ evaluation uses a
larger model applied only to a daily quality-control sample, keeping
per-query cost low. Listing features are pre-cached and reused across daily
runs; booking session pair sampling is $O(n)$ in the session log with no
expensive preprocessing. End-to-end latency from session sampling to database
insertion is under two hours for 10,000 queries on standard data processing
infrastructure.}

\textbf{Daily Generation Volume}: The production pipeline refreshes daily,
generating approximately 10,000 synthetic queries per day from $\sim$500 seed queries---a
20$\times$ amplification factor relative to the survey seed set---sourced from
approximately 5,000 contrastive pairs and using an 80/20 mix of seed-guided and variety prompts.

\subsection{Quality Control}
\label{sec:quality_control}

Each daily run computes quality metrics before queries enter the training
and evaluation databases:

\textbf{Distributional QC}: KL divergence between the current batch's query
length distribution and the seed distribution is computed. Batches exceeding a
pre-defined threshold are flagged for manual review and excluded from
downstream use.

\textbf{Attribute Coverage QC}: The fraction of queries with zero recognized
attributes is tracked against a pre-defined upper bound. Out-of-control batches
trigger alerts and a prompt revision cycle.

\textbf{Label Consistency QC}: For contrastive labels, a random sample is
re-evaluated by the VJ. Batches where VJ label agreement drops below a pre-defined
threshold are quarantined.

\textbf{Deduplication}: Queries within a cosine distance threshold of any existing
query in the database are excluded to maintain diversity.

Dashboard monitoring tracks all metrics daily, combining offline prototype
analysis with online deployment monitoring.

\subsection{Practical Lessons}

\textbf{Start Simple.} Our initial implementation used only listing titles and
descriptions. While limited, this unblocked downstream development
immediately---teams that wait for perfect synthetic data may find themselves
perpetually blocked.

\textbf{Difficulty Calibration.} Uniformly high accuracy provides no signal for
model improvement. Controlled difficulty tiers---achieved through leakage
guardrails, difficulty buckets, and hierarchical pair sampling with controlled
similarity---transform evaluation into a tool that can distinguish incremental
improvements.

\textbf{Model Selection.} Smaller models suffice for bulk generation; larger
models are reserved for quality validation (VJ evaluation). Daily refresh
ensures queries remain current and enables continuous monitoring of generation
drift.

\subsection{Cold-to-Warm Start Transition}
\label{sec:warm_start}

The CASTLE framework supports continuous refinement as real user data gradually becomes available.

\paragraph{Cold-Start (pre-launch).} Approximately 500 curated survey seeds
serve as linguistic templates. Generation uses the full seed-guided pipeline.

\paragraph{Early Warm Start (first weeks post-launch).} Real user queries are
collected from production traffic using stratified sampling by query length
bucket (1--2 words, 3--5 words, 6--8 words, 9+ words) to preserve
distributional coverage. The seed pool is augmented with up to 200 real queries
per week. A KL divergence threshold on real-to-synthetic alignment gates the
reduction of synthetic generation volume.

\paragraph{Mature Warm Start.} As real query volume grows, seed data is
periodically refreshed with a sample of recent production queries. The synthetic
pipeline shifts from bootstrapping to augmenting real data in underrepresented
intent categories; the synthetic-to-real training ratio begins at 10:1 and
decreases as real data accumulates.

\paragraph{Beyond Cold-Start: Targeting the Long Tail.} Real user traffic is
inherently biased toward \emph{head} queries---popular destinations, common
amenity combinations, and mainstream trip types. Tail queries (unusual property
types, niche amenity combinations, edge-case intents) are rare in organic data
but disproportionately important for model robustness. Synthetic generation can
deliberately oversample the tail by conditioning on underrepresented attribute
combinations or incorporating review-derived niche vocabulary. This lifecycle
utility is explored further in Section~\ref{sec:when_synthetic}.

\section{Discussion}

\subsection{When Synthetic Data Works}
\label{sec:when_synthetic}

Synthetic data proves most effective when: (1)~evaluation metrics are
well-defined (pairwise ranking accuracy); (2)~teams need to iterate offline
before real traffic exists; (3)~cold-start requires bootstrap signal; and
(4)~real traffic skews toward head queries and leaves the long tail
underserved (see Section~\ref{sec:warm_start}). The key insight is that
synthetic data need not be perfect---it needs only to be informative enough to
guide development in productive directions.

\subsection{Limitations and Future Directions}
\label{sec:future}

\brandnewblock
\textbf{Input Flexibility.} A practical strength of CASTLE is that its
structured input design is modular: the listing feature set passed to the LLM
is not fixed. In production we use structured attributes (amenities, property
type, location) and review summaries, which already surface experiential and aesthetic
signals---``cozy vibe'', ``stunning mountain views'', ``perfect for
remote work''---that structured attributes alone cannot express. Photo captions
can be incorporated directly alongside these fields, encoding visual style and
ambiance to extend coverage further. Adding these richer signals enables
generation of niche, long-tail queries that reflect subjective user preferences
beyond catalog schema, extending coverage to the very queries that fail under
attribute-only prompting (see failure modes below). Practically, this means
teams can iterate on input richness incrementally: start with structured
attributes to unblock development, then layer in photo signals as they become available.

\brandendblock

\textbf{Failure Modes.} Our generation fails in three systematic patterns
identified through qualitative analysis, illustrated in
Table~\ref{tab:failure_modes}. First, \emph{multi-intent queries}: contrastive
pairs rarely differ on two dimensions simultaneously, so queries like ``romantic
but also great for remote work'' are underrepresented---the LLM cannot identify a
listing pair that clearly separates both intents at once. Second, \emph{very
short queries} (1--2 words): our seed data under-represents these (15.6\%)
relative to real users (34.9\%), and LLM generation conditioned on rich listing
features naturally produces more attribute-rich outputs. Third, \emph{vibe /
experiential queries}: queries like ``cozy and dreamy'' rely on implicit aesthetic
preferences not captured in structured listing features.
Future work will address these through: targeted seed collection for each failure
mode; unsupervised query clustering to identify underrepresented intent
categories; and feature expansion to include image embeddings for vibe queries.

\begin{table}[t]
\caption{Qualitative failure mode examples. Each row shows
a real user query type that synthetic generation struggles to reproduce, with the
root cause and a mitigation strategy.}
\label{tab:failure_modes}
\small
\rowcolors{2}{gray!10}{white}
\begin{tabular}{p{0.13\columnwidth}p{0.22\columnwidth}p{0.22\columnwidth}p{0.27\columnwidth}}
\toprule
\textbf{Type} & \textbf{Real user query} &
\textbf{Synthetic output} &
\textbf{Root cause \& mitigation} \\
\midrule
Multi-intent &
``romantic but also good for remote work'' &
``romantic getaway with ocean view'' &
Contrastive pairs rarely differ on 2 intents simultaneously; need
multi-attribute pair sampling \\
\midrule
Very short &
``cabin'' &
``cozy mountain cabin near ski resort'' &
LLM over-specifies when conditioned on rich features; seed
under-represents 1--2 word queries (15.6\% vs.\ 34.9\%); mitigation:
length-stratified seed sampling to boost short-query representation \\
\midrule
Vibe / experiential &
``dreamy and cozy'' &
``charming cottage with fireplace and garden'' &
Aesthetic vibes not in structured features; mitigation: incorporate visual
features or style tags \\
\bottomrule
\end{tabular}
\end{table}

\textbf{Distribution Shift.} Distribution shift between synthetic and real
queries is inherent---we view synthetic data as a bridge rather than a permanent
replacement. Section~\ref{sec:warm_start} describes our iterative refinement
loop that closes the gap over time. Future work includes a longitudinal study
of how generalization evolves as the seed pool transitions from survey to real
queries, and a formal analysis of scaling under distribution shift (i.e., how
the 1K--5K quality plateau in Table~\ref{tab:scaling} shifts as synthetic data
fidelity improves).

\textbf{Template-Entity Independence.} Our decomposition assumes query
templates and listing entities are independently sampled, but in practice query
style likely correlates with attribute type (``romantic'' queries emphasize
different attributes than ``office'' queries). Future work could model $P(e|t)$
to capture these correlations.

\textbf{Fairness.} Our contrastive pair sampling from booking sessions may
over-represent high-conversion listings, potentially introducing bias against
listings with fewer bookings (e.g., new listings, listings in less-searched
markets). Generated queries may therefore under-represent the search intents
that would surface these listings. Future work will explore fairness-aware
sampling strategies that ensure coverage across listing vintage, price tier, and
geographic market. We acknowledge this as an open limitation of our current
approach.

\textbf{Multi-Turn Conversational Search.} CASTLE extends naturally to
multi-turn conversational search. We envision two conversation categories:
\emph{Refinement} (iterative narrowing, pivoting, or negation of a search) and
\emph{Follow-up Q\&A} (setwise comparisons, factual questions, bookability
queries over a retrieved set). Turn 1 is grounded in real user seed queries;
subsequent turns are generated conditioned on full conversation history.
Cold-start is even more acute here---turn-two and turn-three training examples
are scarce in any new deployment. CASTLE's contrastive component extends
directly: listing-dependent generation grounds Q\&A responses in verifiable
attributes, preventing hallucination in multi-turn contexts. Example
conversation flows are shown below; Turn~1 is seeded from a real user query
and subsequent turns are conditioned on conversation history.

\textbf{Refinement} — user narrows or pivots the search across turns.
\begin{center}
\fbox{\parbox{0.9\columnwidth}{\small\ttfamily
Turn 1 (User): cozy cabin near ski resort\\
Turn 2 (User): under \$200 a night\\
Turn 3 (User): actually needs to sleep 6 people\\
Turn 4 (User): no hot tubs, prefer something more rustic
}}
\end{center}

\textbf{Q\&A} — user asks information-seeking questions over a retrieved listing set.
\begin{center}
\fbox{\parbox{0.9\columnwidth}{\small\ttfamily
Turn 1 (User): beach house with pool in Malibu\\
Turn 2 (User): does it have a private pool or is it shared?\\
Turn 3 (User): which of these two listings is closer to the beach?\\
Turn 4 (User): can I book this for New Year's week?
}}
\end{center}

Longer-horizon directions include a learned user simulator and multimodal turn
generation grounded in visual listing signals.

\section{Conclusion}

We present an LLM-based framework for generating synthetic search queries and labels to
address the cold-start challenge in natural language search. We combine
contrastive listing pairs (grounding queries in real platform features) with
seed queries from structured user research (ensuring realistic linguistic
patterns), developing three prompt variants to balance realism and diversity.
For label generation, we introduce contrastive generation that produces labels
by construction and Virtual Judge labeling for broader coverage.

Compared against three baselines---contrastive-only,
InPars-style, and Promptagator---our seed-guided approach achieves a
9.2$\times$ improvement in query length KL divergence (1.01 vs.\ 9.51/9.33) and
the lowest attribute-type KL divergence (0.08), outperforming even the survey
seed queries themselves (0.09). A human evaluation on 200 triplets confirms
label quality at 91--93\% annotator agreement. We characterize three systematic
failure modes, provide an empirical scaling curve showing quality dominates
volume at this scale, and document production deployment including LLM
selection, quality control, and the warm-start transition protocol.

Open directions include failure-mode-targeted seed collection, fairness-aware
pair sampling, the transition from topicality accuracy to bookability-based
NDCG as real user data accumulates, and extension to multi-turn conversational
search---where CASTLE's contrastive grounding enables listing-dependent
generation of refinement and Q\&A turns conditioned on full conversation
history.

\begin{acks}
We thank the AI Platform---Relevance \& Personalization team at Airbnb for valuable discussions.
\end{acks}

\section*{GenAI Usage Disclosure}

In accordance with CIKM 2026 submission requirements, we disclose the
following uses of generative AI tools in this work:

\textbf{Research}: Large language models are central to
the research methodology and are described in detail in the paper.

\textbf{Writing}: Generative AI tools were used for typographic proofreading and copy-editing assistance. All scientific content, analysis, and conclusions are the authors' own.

\textbf{Code and Data}: LLM API calls were used to generate synthetic queries
and relevance labels as described throughout the paper. No AI tools were used
for data collection or analysis code outside of the described methodology.

\bibliographystyle{ACM-Reference-Format}
\bibliography{references}

\end{document}